# Highlighting relations between Wave-particle duality, Uncertainty principle, Phase space and Microstates

**Ravo Tokiniaina Ranaivoson[1], Voriraza Séraphin Hejesoa[2], Raoelina Andriambololona[3], Nirina Gilbert Rasolofoson[4], Hanitriarivo Rakotoson[5], Jacqueline Rabesahala Raoelina Andriambololona[6], Lala Rarivomanantsoa[7], Naivo Rabesiranana[8]**

*tokhiniaina@gmail.com[1], epistemyandchristian@gmail.com[2], raoelina.andriambololona@gmail.com[3], jacquelineraoelina@hotmail.com[3,6], nirizi.rasolofoson@gmail.com[4], infotsara@gmail.com[5], lala.rarivomanantsoa@gmail.com[7]*

[1,3,4,5,8] *Institut National des Sciences et Techniques Nucléaires (INSTN- Madagascar)*
BP 3907 Antananarivo 101, Madagascar, *instn@moov.mg*

[2,6,7] *Department of Philosophy, University of Antananarivo, Madagascar*
Antananarivo 101, Madagascar

[1, 3, 5,6] TWAS *Madagascar Chapter, Malagasy Academy*
BP 4279 Antananarivo 101, Madagascar

**Abstract:** Wave-particle duality is often considered as the modern answer to the problem of the nature of light after more than 2000 years of questioning. It is also the answer given by quantum physics concerning the nature of matter particles and any other radiations. The main objective of this work is to analyze the relations that are existing between this concept of wave-particle duality, the uncertainty principle and the concepts of phase space and microstates considered in statistical mechanics. It is mainly highlighted that while the concepts of phase space and microstates were already introduced in classical physics before the discovery of the wave-particle duality, a correct understanding of them cannot be achieved without the use of the concept of quantum phase space and phase space representation of quantum mechanics which are directly related to the uncertainty principle. The possibility of using these concepts of quantum phase space and phase space representations of quantum mechanics to help in a deeper description of the wave-particle duality and in the study of some current issues related to foundational problems of quantum mechanics like quantum decoherence and the measurement problem is also discussed.

# 1-Introduction

What is light? This question which may look very simple has survived for more than 2000 years. The interaction between matter and light and their nature has been a major quest for physicists and philosophers since at least Democritus (460-370 BC). Gradually, the theory of light split in two main parts until Planck (1858–1947) and Einstein (1879–1955) works [1-5]. These two parts are corpuscular theory and wave theory. Einstein can be considered as the first who introduced the concept of wave-particle duality (for light) when he gave its corpuscular explanation of the photoelectric effects [3-5] while the wave theory of light was already rigorously established within the Maxwell's theory of electromagnetic waves and its experimental proofs [6]. However, it is well known that the explanation given by Einstein concerning photoelectric effect is based on the results that Planck obtained previously when he tried to solve the famous blackbody radiation problem [1-5]. The extension of the concept of wave-particle duality from the case of light to matter particles by Louis de Broglie (1892-1987) and the deepening and formalization of this idea through the works of Bohr (1885-1962), Heisenberg (1901-1976), Born (1882-1970), Jordan (1902-1980), Pauli (1900-1958), Schrodinger (1887-1961), Dirac (1902-1984), Von Neumann (1903-1957) and many others lead physicists to the foundation of quantum mechanics [4-5,7-8]. Quantum mechanics has successfully explained many experimental results and phenomena that are incomprehensible within the framework of classical physics. However, it is known that there are until now some foundational issues like the famous measurement problem which is often related to the concept of quantum decoherence and the existence of various interpretations of quantum mechanics [9-12]. After the birth of quantum mechanics, the next major step in the evolution of scientific understanding and description of radiations and matter was the advent of relativistic quantum physics which is often associated with Quantum Field Theory [13-16]. The main example of successful Quantum Field Theory is the current Standard Model of Particle Physics [17-19]. Concerning particularly the deepening of the study of the quantum properties of light, with the concept of photons and in relation to technological applications such as lasers, one should also mention the development of quantum optics in particular since the 1950s and 1960s [20-22].

The development of quantum mechanics has permitted to increase significantly our knowledge about microscopic constituents of matter and radiations like light. However, the in-depth study of the macroscopic states of matter (like solid, liquid and gas) and the establishment of a link between their properties and the behavior of their microscopic constituents need the use of statistical mechanics. It is also worth pointing out that even the description of the Blackbody radiation, which was at the origin of quantum mechanics, needed tools from statistical mechanics. Historically, statistical mechanics existed before the development of quantum mechanics. Its foundation is mainly attributed to Boltzmann (1844-1906), Maxwell (1831-1879) and Gibbs (1839-1903) [23-25]. Classical statistical mechanics attempted to establish the link between the mechanical states (called microstates) of the microscopic particles which were supposed to compose a matter sample and the macroscopic thermodynamic state (called macrostate) of this sample. As quantum mechanics didn't exist in the early days of statistical mechanics, physicists have tried to link microstates with the concept of phase space considered in classical Hamiltonian mechanics which was one of the most advanced formulations of classical mechanics. However as we will point out through this work, the only rigorous and correct way to define microstates is to identify them with quantum states. It will also be highlighted through this paper that the concepts of microstates and phase space are deeply linked with the wave-particle duality and the quantum uncertainty principle.

A description of the wave particle duality and its relation with the uncertainty principle is perfomed in the next section (section 2) with also a brief description of the concepts of phase space, microstates and macrostates. Then, detailed presentations of the concepts of quantum phase space and phase space representations of quantum mechanics are given in the sections 3 and 4 respectively. These concepts are used to build the adequate framework which permits to highlight correctly the relation between the wave particle duality, uncertainty principle, phase space and microstates. The contents of the paper are completed with the discussions and conclusions in the section 5. Bold letters are used to represent quantum operators while normal letters correspond to eigenvalues or non-operator quantities.

## 2- Wave particle duality, uncertainty principle, phase space and microstates

### 2.1 Wave particle duality, wavefunctions and uncertainty principle

The concept of wave-particle duality comes naturally, for the case of light and other electromagnetic radiations, from the works of Planck and Einstein [1-5]. Louis de Broglie generalized this concept by extending it in the case of negatons and other matter particles [4-5]. He postulated that like electromagnetic waves are associated to photons, there should be waves associated to matter particles like negatons. The relations between the particle and the associated wave can be summarized by the famous Planck-Einstein-De Broglie relations which formalize the wave particle duality: to a particle with energy $\varepsilon$ and linear momentum $\vec{p}$ is associated a wave with frequency $\nu$, or angular frequency $\omega = 2\pi\nu$, and wave vector $\vec{k}$ so that we have the relations ($\hbar = h/2\pi$ is the Planck reduced constant) [4-5]

$$\begin{cases} \varepsilon = h\nu = \hbar\omega \\ \vec{p} = \hbar\vec{k} \end{cases} \quad (1)$$

Many experimental studies has been conducted to test the De Broglie hypothesis and the results of these experiences are in agreement with the relations in (1) [4-5, 26-28].

Following the wave-particle duality, to a particle should be associated a wave. The wave nature is described in elementary quantum mechanics by the introduction of a wavefunction $\psi(\vec{r})$. In the framework of the probabilistic interpretation of quantum mechanics, the wavefunction $\psi(\vec{r})$ is related to the density of probability of finding the particle around a particular position $\vec{r}$ at the instant $t$: more precisely, this density is considered as equal to the square $|\psi(\vec{r})|^2$ of the modulus of $\psi(\vec{r})$[4]. The wavefunctions were introduced by Schrödinger in 1926 [4-5, 29] but the origin of the probabilistic interpretation is attributed to Max Born [5].

The wavefunction $\psi(\vec{r})$ is more precisely called coordinate wavefunction or wavefunction in coordinate representation because it describes the probability associated to the coordinates of the particle in ordinary space. To describe the probability density associated to the linear momentum $\vec{p}$ of the particle, a momentum wavefunction or wavefunction in momentum representation, denoted $\tilde{\psi}(\vec{p})$ is introduced. Like in the case of coordinate wavefunction, the square $|\tilde{\psi}(\vec{p})|^2$ of the modulus of the momentum wavefunction $\tilde{\psi}(\vec{p})$ is considered to be the probability density associated to $\vec{p}$.

The wavefunctions $\psi(\vec{r})$ and $\tilde{\psi}(\vec{p})$ can be considered as « the components » of the quantum state vector $|\psi\rangle$ of a particle, respectively in the basis $\{|\vec{r}\rangle\}$ and $\{|\vec{p}\rangle\}$. With $\{|\vec{r}\rangle\}$ the basis of the states space of the particle formed by the eigenstates $\{|\vec{r}\rangle\}$ of the position operator $\vec{r}$ and $\{|\vec{p}\rangle\}$ the basis of the states space of the particle formed by the eigenstates $\{|\vec{p}\rangle\}$ of the momentum operator $\boldsymbol{\vec{p}}$. More explicitly, one has the following relations

$$|\psi\rangle = \iiint |\vec{r}\rangle\langle\vec{r}|\psi\rangle \, d^3\vec{r} = \iiint |\vec{r}\rangle \psi(\vec{r}) d^3\vec{r} \Leftrightarrow \psi(\vec{r}) = \langle\vec{r}|\psi\rangle \quad (2)$$

$$|\psi\rangle = \iiint |\vec{p}\rangle\langle\vec{p}|\psi\rangle \, d^3\vec{p} = \iiint |\vec{p}\rangle \tilde{\psi}(\vec{p}) d^3\vec{p} \Leftrightarrow \tilde{\psi}(\vec{p}) = \langle\vec{p}|\psi\rangle \quad (3)$$

$$\tilde{\psi}(\vec{p}) = \langle\vec{p}|\psi\rangle = \iiint \langle\vec{p}|\vec{r}\rangle\langle\vec{r}|\psi\rangle \, d^3\vec{r} = \frac{1}{(2\pi\hbar)^{3/2}} \iiint e^{-i\frac{\vec{p}\cdot\vec{r}}{\hbar}} \psi(\vec{r}) \, d^3\vec{r} \quad (4)$$

The relation (4) is a Fourier-like transform and it corresponds to a basis change between the basis $\{|\vec{r}\rangle\}$ and $\{|\vec{p}\rangle\}$. One has respectively, in the coordinate representation, for the position operator $\vec{r}$ and momentum operator $\vec{p}$ the following relations [4-5].

$$\begin{cases} \langle\vec{r}|\vec{r}|\psi\rangle = \vec{r}\langle\vec{r}|\psi\rangle = \vec{r}\psi(\vec{r}) \\ \langle\vec{r}|\vec{p}|\psi\rangle = -i\hbar\vec{\nabla}[\langle\vec{r}|\psi\rangle] = -i\hbar\vec{\nabla}[\psi(\vec{r})] \end{cases} \quad (5)$$

Given the probabilistic interpretation of the wavefunctions, the statistical mean values of the components of $\vec{r} = \vec{e}_x x + \vec{e}_y y + \vec{e}_z z$ and $\vec{p} = \vec{e}_x p_x + \vec{e}_y p_y + \vec{e}_z p_z$ and the corresponding statistical variances can be defined like in any probabilistic theory. We have for instance for the mean values $\langle x \rangle, \langle p_x \rangle$ and variances $(\sigma_x)^2, (\sigma_{p_x})^2$ of $x$ and $p_x$

$$\begin{cases} \langle x \rangle = \langle\psi|x|\psi\rangle = \iiint x|\psi(\vec{r})|^2 \, d^3\vec{r} \\ \langle p_x \rangle = \langle\psi|x|\psi\rangle = \iiint p_x|\tilde{\psi}(\vec{p})|^2 \, d^3\vec{p} = \iiint [\psi^*(\vec{r})(-i\hbar\frac{\partial\psi(\vec{r})}{\partial x})]d^3\vec{r} \\ (\sigma_x)^2 = \langle\psi|(x-\langle x\rangle)^2|\psi\rangle = \iiint (x-\langle x\rangle)^2|\psi(\vec{r})|^2 \, d^3\vec{r} \\ (\sigma_{p_x})^2 = \langle\psi|(p-\langle p\rangle)^2|\psi\rangle = \iiint (p_x-\langle p_x\rangle)^2|\tilde{\psi}(\vec{p})|^2 d^3\vec{p} \end{cases} \quad (6)$$

Using Cauchy-Schwarz inequality, it can be deduced directly from the relations (4), (5) and (6) the following inequality [4-5, 30].

$$(\sigma_x)(\sigma_{p_x}) \geq \frac{\hbar}{2} \quad (7)$$

The relation (7) is the mathematical form of the uncertainty principle, i.e. it is the uncertainty relation, for the case of operators $x$ and $p_x$. Analogous relations exist for the other components of the operators $\vec{r}$ and $\vec{p}$. More generally, from the basic postulates of quantum mechanics, it can be deduced that the uncertainty relations exist not only between position and momentum but between any two observables $A$ and $B$ that are not commutative [4-5, 31]. Let $C$ be the commutator of $A$ and $B$ i.e. $AB - BA = C$ and $|\psi\rangle$ a state of a quantum system. We may consider respectively the mean values $\langle A\rangle, \langle B\rangle, \langle C\rangle$ and the statistical variances $(\sigma_A)^2, (\sigma_B)^2$ of the operators $A, B$ and $C$ corresponding to the state $|\psi\rangle$.

$$\begin{cases} \langle A\rangle = \langle\psi|A|\psi\rangle & (\sigma_A)^2 = \langle\psi|(A-\langle A\rangle)^2|\psi\rangle \\ \langle B\rangle = \langle\psi|B|\psi\rangle & (\sigma_B)^2 = \langle\psi|(B-\langle B\rangle)^2|\psi\rangle \\ \langle C\rangle = \langle\psi|C|\psi\rangle \end{cases} \quad (8)$$

From the properties of the states space of the quantum system, which is a Hilbert space, one deduce the Robertson uncertainty relation [4-5, 31]

$$(\sigma_A)(\sigma_B) \geq \frac{1}{2}|\langle C \rangle| \tag{9}$$

The relation (9) is the mathematical form of the uncertainty principle applicable to any pair of quantum operators **A** and **B**. According to this principle, there is a limit to the precision with which the values of the observables pair can be simultaneously known.

## 2.2 Phase space and uncertainty principle

The concept of phase space is known in the framework of classical mechanics as closely related to Hamiltonian mechanics and statistical mechanics [23-25, 32-34]. The elementary example of classical phase space is the phase space associated to a particle performing monodimensional motion with a coordinate $x$ and momentum $p$ : for this case, the phase space is just defined as the set $\{(p, x)\}$ of the possible values of the pair $(p, x)$. The topological dimension of this phase space is equal to 2 and it can be geometrically identified with a plan. More generally, for a system with a degree of freedom equal to $D$, the classical phase space is defined as the set of possible values of the momentum-coordinate pairs $(p_l, x_l)$ with $l = 1, 2, \dots, D$ : the topological dimension of this phase space is equal to $2D$. From the deterministic point of view of classical physics, it is conceptually possible to know simultaneously the exact instantaneous values $p_1(t), x_1(t), \dots p_D(t), x_D(t)$ of the $2D$ momenta and coordinates at any instant $t$. It means that it is possible to associate, at any time, to a classical state of the system a single point in its classical phase space. It follows that from the point of view of classical mechanics, the phase space can be considered as the set of possible states of a system: each point on this phase space corresponds to one and one possible classical state of the system.

It is not easy to extend the definition of phase space from classical physics to quantum physics because of the uncertainty relations which are consequences of the Canonical Commutation Relations (CCRs). These CCRs can be put in the form

$$\begin{cases} [p_l, x_k] = p_l x_k - x_k p_l = -i\hbar \delta_{lk} = \begin{cases} -i\hbar \ si \ l = k \\ 0 \ \ si \ l \neq k \end{cases} \\ [p_l, p_k] = p_l p_k - p_k p_l = 0 \\ [x_l, x_k] = x_l x_k - x_k x_l = 0 \end{cases} \tag{10}$$

Then, given the relation (9), it follows from (10) that one has for every pairs of quantum observables $(\boldsymbol{p}_l, \boldsymbol{x}_l)$ an uncertainty relation analogous to (7)

$$(\sigma_{p_l})(\sigma_{x_l}) \geq \frac{\hbar}{2} \tag{11}$$

According to the relation (11), one cannot consider rigorously simultaneous exact values of $p_l$ and $x_l$. It follows that the extension of the concept of phase space from classical mechanics to quantum mechanics is not trivial. In the section 3, we will look at a solution to this issue.

## 2.3 Microstates, phase space and macrostates

In the framework of statistical mechanics, the main objective is to relate the microscopic states of particles composing a system to the macroscopic properties of the system which correspond to its macrostates. In the framework of classical physics, the microscopic states of particles are identified as their mechanical states while the macroscopic states of the system are identified with its thermodynamic states. The mechanical states which correspond conceptually to details on the "microscopic motions" of each particles are called microstates while the thermodynamic states which describe a "macroscopic behavior" of the system are called macrostates. To a single macrostate may correspond many possible microstates. In the microcanonical approach of statistical mechanics, for instance, the relation between a macrostate and the corresponding microstates can be described with the introduction of the Boltzmann entropy $S$ [25]:

$$S = k_B ln(\Omega) \qquad (12)$$

in which $k_B$ is the Boltzmann constant and $\Omega$ is the number of possible microstates associated to the macrostate that is considered. Any thermodynamic property of the system can be deduced if the explicit expression of the entropy $S$ is known. It is then necessary to "count" the microstates associated to the considered macrostate i.e. to find $\Omega$. It can be done by considering the hypervolume $\mathcal{V}$ in phase space which corresponds to the set of the microstates associated to the considered macrostate and to divide it by the "elementary hypervolume $v$" of a microstate

$$\Omega = \frac{\mathcal{V}}{v} \qquad (13)$$

There is no known classical method which permits to determine the value of the elementary hypervolume $v$ corresponding to a microstate. Then even in the framework of what is called "classical statistical mechanics'', it is just later commonly accepted that $v$ should be taken as equal to $(h)^D$, when the dimension of the classical phase space of the system which is considered is equal to $2D$ ($h$ being the Planck constant). This practice means that the formulation of "classical statistical mechanics" need intrinsically at its core the introduction of the Planck constant $h$. This fact may lead to expect that there is possibly a natural relation between the concept of microstates and quantum mechanics.

From the deterministic point of view of classical mechanics which is discussed in the beginning of the section 2.2 related to classical phase space, it may seem reasonable to identify a microstate of a system with a single point in its classical phase space. However the fact that a real microstate should have a "nonzero hypervolume", $v = (h)^D$, is in contradiction with this point of view. It follows that a microstate cannot be identified with a determined mechanical state of a system as it is considered in classical mechanics. In other words the concept of phase space in classical mechanics is not rigorously compatible with the concept of microstate as it is considered in statistical mechanics. As we will see in the section 3, the concept of quantum phase space suggests an adequate solution to this problem.

The concept of microstates is always at the core of statistical mechanics. The identification of a microstate with a quantum state is considered as common in the framework of quantum statistical mechanics. However, we will show explicitly in the next paragraph that "any microstates" should be always considered as a quantum state even the microstates considered in "classical statistical mechanics". For the case of the macrostates, they are classically commonly identified with the thermodynamic states. In the case of quantum statistical mechanics, one can have a more explicit description of macrostates by using a formalism of quantum mechanics, introduced by John von Neumann, which is based on the representation of states with density operators. Some examples of interesting studies related to density operators and their uses can be for instance found in the references [33-42]. A density operator $\boldsymbol{\rho}$ is an Hermitian operator which is positive semi-definite and has a trace equal to 1. As an operator, $\boldsymbol{\rho}$ is acting on the Hilbert (states) space of the system which is considered. Two main cases can be considered:

- For an ordinary quantum state $|\psi\rangle$, called a pure state, the corresponding density operator $\boldsymbol{\rho}$ is given by the relation

$$\boldsymbol{\rho} = |\psi\rangle\langle\psi| \qquad (14)$$

- For a mixed state which corresponds to a statistical ensemble, the density operator is of the form

$$\boldsymbol{\rho} = \sum_I p_I |\psi_I\rangle\langle\psi_I| \qquad (15)$$

In the framework of quantum statistical mechanics, the density operator in the relation (14) corresponds to a microstate while the relation (15) is associated to a macrostate. The coefficients $p_I$ in (15) are positive numbers. Each of them can be interpreted as the "classical-like probability" of the system to be in a particular microstate (pure state) $|\psi_I\rangle$. Each $p_I$ is called a "classical like probability" because the statistical combination described in the relation (15) corresponds to a classical-like incoherent combination of the states $|\psi_I\rangle$ which is fundamentally different from coherent quantum superposition of states. Besides, the family $\{|\psi_I\rangle\}$ is not necessarily a basis of the Hilbert space of the system and the pure states $|\psi_I\rangle$ are not necessarily orthogonal to each other. The decomposition (15) of the operator $\boldsymbol{\rho}$ is also not unique: there are in general an infinite number of statistical ensemble which can correspond to the same density operator.

Given the density operator $\boldsymbol{\rho}$ of a system, the mean value $\langle\boldsymbol{A}\rangle$ of an observable $\boldsymbol{A}$ is given by the relation

$$\langle\boldsymbol{A}\rangle = Tr(\boldsymbol{\rho}\boldsymbol{A}) \qquad (16)$$

in which $Tr$ refers to operator trace. If the density operator $\boldsymbol{\rho}$ corresponds to a macrostate, the relation (16) may give directly the value of a macroscopic quantity. The mean value $\langle\boldsymbol{H}\rangle$ of the Hamiltonian operator, for instance, corresponds directly to the value of the thermodynamic internal energy $U$ of a system: $U = \langle\boldsymbol{H}\rangle = tr(\boldsymbol{\rho}\boldsymbol{H})$ [33-34].

The elementary time evolution equation of the density operator $\boldsymbol{\rho}$ is the Liouville-Von Neumann equation

$$i\hbar \frac{d\boldsymbol{\rho}}{dt} = [\boldsymbol{H}, \boldsymbol{\rho}] \qquad (17)$$

This equation is valid for both pure state and mixed states when one deals with unitary evolution (ordinary quantum evolution). However, this equation is non-longer sufficient when one deals with more general cases like in the study of open quantum system, quantum decoherence and quantum thermodynamics. In this cases, more accurate equations are used [38-42].

## 3- Joint momenta-coordinates states and quantum phase space

### 3.1 Joint momentum-position states: case of one momentum-position pair

Because of the uncertainty relations, it is not easy to conciliate the concept of phase space with quantum mechanics. There is a formulation of quantum mechanics, known as the phase space formulation [43-50], which corresponds to an approach that deals with this issue. This formulation is based on the use of a quasiprobability distribution introduced in the work [43] of E.P. Wigner (1902-1995). Historically, the first main contributors in the development of this formulation was Wigner, Weyl (1885-1955), Groenewold (1910-1996) and Moyal (1910-1998). This phase space formulation of quantum mechanics has led to interesting results and is now accepted as being one of the possible alternative to formulate quantum mechanics. However, the Wigner quasiprobability distribution, which should be interpreted as a density of probability of localization in phase space has some unusual properties. It can, for instance, have negative values. These unusual properties may be explained with the fact that the phase space considered in the formulation is a classical phase space, i.e. the Wigner quasiprobability distribution is a representation of a quantum state in a classical phase space. Another approach, which is based on the concept of quantum phase space was considered through the references [33-34, 51-52]. This approach is based of the introduction of momentum-coordinate joint states. Let us first consider the case of a single pair $(\boldsymbol{p}, \boldsymbol{x}) = (\boldsymbol{p}_1, \boldsymbol{x}_1)$ of a momentum operator $\boldsymbol{p}_1$ and coordinate operator $\boldsymbol{x}_1$. This pair is characterized by the Canonical Commutation Relation (CCR)

$$[\boldsymbol{p}_1, \boldsymbol{x}_1] = \boldsymbol{p}_1\boldsymbol{x}_1 - \boldsymbol{x}_1\boldsymbol{p}_1 = -i\hbar \qquad (18)$$

The form of the CCR in (18) is chosen in view of a multidimensional and relativistic generalization. In this view, we can also highlight distinction between covariants momentum and position operators $\boldsymbol{p}_1$ and $\boldsymbol{x}_1$ (low index) and their contravariants analogous $\boldsymbol{p}^1$ and $\boldsymbol{x}^1$ (high index). The identification of the CCR (18), $[\boldsymbol{p}_1, \boldsymbol{x}_1] = -i\hbar$, as a particular case of the multidimensional and relativistic relation $[\boldsymbol{p}_\mu, \boldsymbol{x}_\nu] = i\hbar\eta_{\mu\nu}$ [33,51] that is considered in the section 3.2 permits to identify the "metric" $\eta = \eta_{11} = -1$. One has the relation [33]

$$\begin{cases} \boldsymbol{p}_1 = \eta_{11}\boldsymbol{p}^1 = -\boldsymbol{p}^1 \Leftrightarrow \boldsymbol{p}^1 = -\boldsymbol{p}_1 \\ \boldsymbol{x}_1 = \eta_{11}\boldsymbol{x}^1 = -\boldsymbol{x}^1 \Leftrightarrow \boldsymbol{x}^1 = -\boldsymbol{x}_1 \end{cases} \qquad (19)$$

For any quantum state $|\psi\rangle$, one may consider the following means values and statistical variance-covariances

$$\begin{cases} \langle p_1 \rangle = \langle\psi|\boldsymbol{p}_1|\psi\rangle = -\langle\psi|\boldsymbol{p}^1|\psi\rangle = -\langle p^1 \rangle \\ \langle x_1 \rangle = \langle\psi|\boldsymbol{x}_1|\psi\rangle = -\langle\psi|\boldsymbol{x}^1|\psi\rangle = -\langle x^1 \rangle \\ \mathcal{P}_{11} = \langle\psi|(\boldsymbol{p}_1 - \langle p_1\rangle)^2|\psi\rangle = -\mathcal{P}_1^1 = \mathcal{P}^{11} \\ \mathcal{X}_{11} = \langle\psi|(\boldsymbol{x}_1 - \langle x_1\rangle)^2|\psi\rangle = -\mathcal{X}_1^1 = \mathcal{X}^{11} \\ \varrho_{11}^{\ltimes} = \langle\psi|(\boldsymbol{p}_1 - \langle p_1\rangle)(\boldsymbol{x}_1 - \langle x_1\rangle)|\psi\rangle \\ \varrho_{11}^{\rtimes} = \langle\psi|(\boldsymbol{x}_1 - \langle x_1\rangle)(\boldsymbol{p}_1 - \langle p_1\rangle)|\psi\rangle \\ \varrho_{11} = \frac{1}{2}(\varrho_{11}^{\ltimes} + \varrho_{11}^{\rtimes}) \end{cases} \qquad (20)$$

Given the CCR (18), the following relation is deduced

$$\varrho_{11} = \frac{1}{2}(\varrho_{11}^{\ltimes} + \varrho_{11}^{\rtimes}) = \varrho_{11}^{\ltimes} + i\frac{\hbar}{2} = \varrho_{11}^{\rtimes} - i\frac{\hbar}{2} \qquad (21)$$

One introduces the momentum-coordinate statistical variance-covariances matrix $\begin{pmatrix} \mathcal{P}_{11} & \varrho_{11} \\ \varrho_{11} & \mathcal{X}_{11} \end{pmatrix}$. Using the property of scalar product on the quantum states space and Cauchy-Schwarz inequality associated to it, it can be shown that the determinant $\begin{vmatrix} \mathcal{P}_{11} & \varrho_{11} \\ \varrho_{11} & \mathcal{X}_{11} \end{vmatrix}$ of this matrix satisfies the following relation [33]

$$\begin{vmatrix} \mathcal{P}_{11} & \varrho_{11} \\ \varrho_{11} & \mathcal{X}_{11} \end{vmatrix} = (\mathcal{P}_{11})(\mathcal{X}_{11}) - (\varrho_{11})^2 \geq \frac{(\hbar)^2}{4} \quad (22)$$

The relation (22) can be considered as an explicit form of the uncertainty relation which is more rigorous than the relation (7). In fact, it includes also the statistical momentum-coordinate statistical covariance $\varrho_{11}$ which is missing in (7). On one hand, according to (22), it is impossible to have a quantum state $|\psi\rangle$ for which one has simultaneous determined values of momentum and coordinate. However, on the other hand, simultaneous values of these variables must be considered within the concept of phase space. To council the concept of phase space with the uncertainty relation (22), one should then consider quantum "momentum-coordinate joint states" that do not violates this uncertainty relation. The best kind of states that satisfy this criterion are states which saturates the relation (22) i.e. states for which one has an equality instead of the inequality. This kind of state, that one may denote $|\langle p_1\rangle, \langle x_1\rangle, \mathcal{P}_{11}, \varrho_{11}\rangle$ will be defined by the means values $\langle p_1 \rangle$ and $\langle x_1 \rangle$ for a given value of the statistical variance-covariances matrix $\begin{pmatrix} \mathcal{P}_{11} & \varrho_{11} \\ \varrho_{11} & \mathcal{X}_{11} \end{pmatrix}$. This kind of states have been identified and considered in the references [33, 51-52]. They corresponds to Gaussian-like wavefunctions. On has for instance for the expression of the coordinate wavefunctions [33, 51]

$$\langle x^1|\langle p_1\rangle, \langle x_1\rangle, \mathcal{P}_{11}, \varrho_{11}\rangle = \left(\frac{1}{2\pi\mathcal{X}_{11}}\right)^{1/4} e^{-\frac{\mathcal{B}_{11}}{\hbar^2}(x^1 - \langle x^1\rangle)^2 - \frac{i}{\hbar}\langle p_1\rangle x^1 + iK} \quad (23)$$

with

$$\mathcal{B}_{11} = -i\frac{\varrho_{11}^\times}{2\mathcal{X}_{11}} = \frac{(\hbar)^2}{4\mathcal{X}_{11}} - i\frac{\hbar\varrho_{11}}{2\mathcal{X}_{11}} = \frac{\mathcal{P}_{11}}{1 + \frac{(\varrho_{11})^2}{4\hbar^2}}\left(1 - i\frac{\varrho_{11}}{2\hbar^2}\right) \quad (24)$$

The relation (23) defines completely a state $|\langle p_1\rangle, \langle x_1\rangle, \mathcal{P}_{11}, \varrho_{11}\rangle$. The parameter $K$ is an arbitrary real number which does not depend on $x^1$ but may depend on $\langle x^1\rangle$ and $\langle p_1\rangle$. It can be shown that a state $|\langle p_1\rangle, \langle x_1\rangle, \mathcal{P}_{11}, \varrho_{11}\rangle$ is an eigenestate of the operator

$$\mathbf{z}_1 = \mathbf{p}_1 - \frac{2i}{\hbar}\mathcal{B}_{11}\mathbf{x}^1 \quad (25)$$

The corresponding eigenvalue equation is

$$\mathbf{z}_1|\langle p_1\rangle, \langle x_1\rangle, \mathcal{P}_{11}, \varrho_{11}\rangle = \left(\langle p_1\rangle - \frac{2i}{\hbar}\mathcal{B}_{11}\langle x^1\rangle\right)|\langle p_1\rangle, \langle x_1\rangle, \mathcal{P}_{11}, \varrho_{11}\rangle = \langle z_1\rangle|\langle z_1\rangle\rangle \quad (26)$$

Following the relation (26), the eigenvalue of the operator $\mathbf{z}_1$ for a state $|\langle p_1\rangle, \langle x_1\rangle, \mathcal{P}_{11}, \varrho_{11}\rangle$ is its mean value $\langle z_1\rangle = \langle p_1\rangle - \frac{2i}{\hbar}\mathcal{B}_{11}\langle x^1\rangle$ itself. So in the last equality in (26), we adopt the notation $|\langle z_1\rangle\rangle = |\langle p_1\rangle, \langle x_1\rangle, \mathcal{P}_{11}, \varrho_{11}\rangle$ as in the references [33, 51].

It can be shown that the statistical variance-covariance matrix $\begin{pmatrix} \mathcal{P}_{11} & \varrho_{11} \\ \varrho_{11} & \mathcal{X}_{11} \end{pmatrix}$ can be put in the form [33]

$$\begin{pmatrix} \mathcal{P}_{11} & \varrho_{11} \\ \varrho_{11} & \mathcal{X}_{11} \end{pmatrix} = \begin{pmatrix} \mathcal{b}_1^1 & 0 \\ c_1^1 & a_1^1 \end{pmatrix}^T \begin{pmatrix} \eta & 0 \\ 0 & \eta \end{pmatrix} \begin{pmatrix} \mathcal{b}_1^1 & 0 \\ c_1^1 & a_1^1 \end{pmatrix} = \begin{pmatrix} \mathcal{b}_1^1 & 0 \\ c_1^1 & a_1^1 \end{pmatrix}^T \begin{pmatrix} -1 & 0 \\ 0 & -1 \end{pmatrix} \begin{pmatrix} \mathcal{b}_1^1 & 0 \\ c_1^1 & a_1^1 \end{pmatrix} \quad (27)$$

with

$$\begin{cases} (a_1^1)^2 = \mathcal{X}_1^1 = -\mathcal{X}_{11} \Rightarrow a_1^1 = i\sqrt{\mathcal{X}_{11}} \\ \mathcal{b}_1^1 = \frac{1}{2a_1^1} = \frac{-i}{2\sqrt{\mathcal{X}_{11}}} \quad c_1^1 = \frac{\varrho_{11}}{a_1^1} = \frac{-i\varrho_{11}}{\sqrt{\mathcal{X}_{11}}} \end{cases} \quad (28)$$

Using the parameter $a_1^1$ in the relations (27) and (28), we may define the following reduced operators

$$\begin{cases} \mathbf{z}_1 = a_1^1(\mathbf{z}_1 - \langle z_1 \rangle) \\ \mathbf{z}_1^\dagger = a_1^{1*}(\mathbf{z}_1^\dagger - \langle z_1 \rangle^*) \end{cases} \quad [\mathbf{z}_1, \mathbf{z}_1^\dagger] = \mathbf{z}_1 \mathbf{z}_1^\dagger - \mathbf{z}_1^\dagger \mathbf{z}_1 = 1 \quad (29)$$

According to their commutation relation, the operators $\mathbf{z}_1$ and $\mathbf{z}_1^\dagger$ have the properties of ladder operators analogous to those considered in the theory of quantum harmonic oscillator. Using these operators, one can identify other kind of momentum-coordinate joint states denoted $|n_1, \langle z_1 \rangle\rangle$, with $n_1$ a positive integer. The relations between the states $|n_1, \langle z_1 \rangle\rangle$ and the states $|\langle z_1 \rangle\rangle$ are analogous to the relations between the excited and ground states of a harmonic oscillator. More explicitly, one has the following relations

$$\begin{cases} |n_1, \langle z_1 \rangle\rangle = \frac{(\mathbf{z}_1^\dagger)^{n_1}}{\sqrt{n_1!}} |\langle z_1 \rangle\rangle \\ \mathbf{z}_1 |n_1, \langle z_1 \rangle\rangle = \sqrt{n_1} |n_1 - 1, \langle z_1 \rangle\rangle \\ \mathbf{z}_1^\dagger |n_1, \langle z_1 \rangle\rangle = \sqrt{n_1 + 1} |n_1 + 1, \langle z_1 \rangle\rangle \end{cases} \quad (30)$$

It can be shown easily that the states $|n_1, \langle z_1 \rangle\rangle$ are eigenstates of the operator $\aleph = \mathbf{z}_1^\dagger \mathbf{z}_1$ with the eigenvalue equation

$$\aleph |n_1, \langle z_1 \rangle\rangle = n_1 |n_1, \langle z_1 \rangle\rangle \quad (31)$$

### 3.2 Quantum Phase Space: case of one momentum-coordinate pair

As in [33], the quantum phase space can be defined as the set $\{(\langle p_1 \rangle, \langle x_1 \rangle)\}$ of the possible values of $\langle p_1 \rangle$ and $\langle x_1 \rangle$ for a given value of the statistical variance-covariance matrix $\begin{pmatrix} \mathcal{P}_{11} & \varrho_{11} \\ \varrho_{11} & \mathcal{X}_{11} \end{pmatrix}$ or equivalently as the set $\{\langle z_1 \rangle\}$ of possible values of $\langle z_1 \rangle$. The structure of this quantum phase space is parameterized by this variance-covariance matrix. It was shown in [33-34] that the variance-covariances, i.e. the elements of this matrix, can be related directly with thermodynamic variables. It follows that the structure of the quantum phase space itself is directly related to thermodynamic.

When the state of the considered system is exactly a state $|\langle z_1 \rangle'\rangle$, one can associate with it the point $\langle z_1 \rangle'$ in the quantum phase space. However, it should be remarked that two states $|\langle z_1 \rangle\rangle$ and $|\langle z_1\rangle'\rangle$ are not orthogonal [33] i.e. one has $\langle\langle z_1\rangle|\langle z_1\rangle'\rangle \neq 0$. It follows that the density of probability $|\langle\langle z_1\rangle|\langle z_1\rangle'\rangle|^2$ to find the system in the state $|\langle z_1\rangle\rangle$ if its state is exactly $|\langle z_1\rangle'\rangle$ (or vice versa) is not zero. Then, if the state of the system is represented by the point $\langle z_1\rangle'$ in the quantum phase space, it is also probable to find it in another state corresponding to the point $\langle z_1\rangle \neq \langle z_1\rangle'$ if a simultaneous measurement of momentum and coordinate is performed. More generally, if the state of the system is a general state $|\psi\rangle$, one can always define the density of probability $|\langle\langle z_1\rangle|\psi\rangle|^2$ to find it in the state $|\langle z_1\rangle\rangle$ represented by the point $\langle z_1\rangle$ in the quantum phase space. The quantity $\widetilde{\psi}(\langle z_1\rangle) = \langle\langle z_1\rangle|\psi\rangle$ can then be considered as a phase space wavefunction or wavefunction in a "phase space representation". This concept of "phase space representation" will be deepened in the next section (section 4).

It can be remarked that the concepts of joint momentum-position states and quantum phase space that was considered could be linked with the concept of pointer states that are studied in the framework of quantum decoherence theory [12, 53-55]. Pointer states are defined as the most stable states that a system tends to occupy through the decoherence phenomena. In other words, these quantum states are the most similar to the classical states. Regardless of the initial quantum states of the system, decoherence phenomena produced by interaction with the environment will make this initial state to tend to transform into one of the pointer states through a kind of an "environment superselection" process [12]. These pointer states are the eigenstates of observables with values that are "continuously checked" i.e. "measured" by the environment. So, for instance, when interactions with the environment depend explicitly on the coordinates observables of a particle, the pointer states will be the eigenstates of this observable i.e. the particles coordinates [12]. However, if a deeper analysis is performed, the natural interactions of particle with their environments most often result in simultaneous measurements of momenta and coordinates observables (via collisions or scattering phenomena for instance). This is in agreement with the fact that classical mechanics intuitively considers these two kind of quantities (momenta and coordinates) as central and fundamental in defining the phase space. Following this point of view, the "pointer states" are expected to be "joint momenta-coordinates states" that are the most similar to the classical states. In other words, these pointer states should be the joint momenta-coordinates states that corresponds to the saturation of the uncertainty principle i.e. the states $|\langle z_1\rangle\rangle$. Thus, the quantum phase space could also be considered as being the set of possible joint momenta-coordinates states that particles tend to occupy under the influence of their natural environments.

If we analyze our approach, the uncertainty relation (22), the introduction of the "momenta-coordinates joint states" (25) as well as the definition of the quantum phase space which was deduced are consequences of (i.e. are deduced from) the Canonical Commutation Relation (18). For the multidimensional and relativistic generalization of the concept of quantum phase space, it is therefore necessary to search firstly for the generalization of this CCR.

### 3.3 Multidimensional and relativistic Canonical Commutation Relations (CCRs)

Let us consider the case of a system with $D$ degrees of freedom and which is described with $D$ momenta-coordinates observables pairs $(\boldsymbol{p}_l, \boldsymbol{x}_l)$ with $l = 1,2,\dots,D$. The elements of these pairs satisfy, by definitions, the Canonical Commutation Relations (CCRs) which are

$$\begin{cases} [\,p_l, x_k] = p_l x_k - x_k p_l = -i\hbar \delta_{lk} = \begin{cases} -i\hbar \; if\; l = k \\ 0 \;\; if\; l \neq k \end{cases} \\ [\,p_l, p_k] = p_l p_k - p_k p_l = 0 \\ [\,x_l, x_k] = x_l x_k - x_k x_l = 0 \end{cases} \qquad (32)$$

A more general case is the case of a relativistic quantum system. According to the relativity theory, space and time coordinates should be put on an equal footing. Besides, these spacetime coordinates are mixed when considering Lorentz transformation [13, 33, 56]. However the concept of time operator is known to be a subject of long-standing debate in quantum mechanics [51, 57-64]. There is between energy and time an uncertainty relation which is similar to the one existing between a spatial momentum and the associated coordinate but the interpretation of the time-energy uncertainty relation is not trivial [57-58].

Besides, according to a Pauli theorem, a quantum time operator cannot exists and time should be only considered a simple parameter in quantum mechanics [51, 58-64]. However, various works have already shown that this Pauli theorem had flaws and limits and that it is then possible to introduce a quantum time observable [51, 60-64]. It was also already shown in our previous works [33, 51, 65] that the introduction of a time operator analogous to spatial coordinates operators can lead to some interesting consequences especially in the framework of particle physics. The existence of this kind of time operator permits also to satisfy the symmetry between space and time. So we admit, as in [33, 51, 65] the existence of a time operator. For a single particle with a quadridimensional spacetime, for instance, one has a four pairs of energy-momenta and spacetime coordinates operators $(p_\mu, x_\mu)$ with $\mu = 0,1,2,3$. In which $(p_0, x_0) = (\frac{\varepsilon}{c}, ct)$ with $\varepsilon$ and $t$, respectively the energy and time operators and $c$ the speed of light in vacuum. And the $(p_l, x_l)$ for $l = 1,2,3$ correspond to the three pairs of spatial momenta-coordinates observables. The corresponding CCR can be written in the following form [33, 51]

$$\begin{cases} [\,p_\mu, x_\nu] = p_\mu x_\nu - x_\nu p_\mu = i\hbar \eta_{\mu\nu} = \begin{cases} i\hbar & if\; \mu = \nu = 0 \\ -i\hbar & if\; \mu = \nu = l = 1,2,3 \\ 0 & if\; \mu \neq \nu \end{cases} \\ [\,p_\mu, p_\nu] = p_\mu p_\nu - p_\nu p_\mu = 0 \\ [\,x_l, x_k] = x_\mu x_\nu - x_\nu x_\mu = 0 \end{cases} \qquad (33)$$

The quantity $\eta_{\mu\nu}$ in (33) correspond to the signature of the spacetime that is considered, which is $(+,-,-,-) = (1,3)$, in the present case. More generally, one may consider general multidimensional and relativistic CCRs which correspond to a general signature $(D_+, D_-)$ with a number of pair $(p_\mu, x_\mu)$ equal to $D_+ + D_- = D$ i.e. $\mu = 0,1,2,\ldots,D-1$. The general form of the corresponding CCRs is

$$\begin{cases} [\,p_\mu, x_\nu] = p_\mu x_\nu - x_\nu p_\mu = i\hbar \eta_{\mu\nu} \\ [\,p_\mu, p_\nu] = p_\mu p_\nu - p_\nu p_\mu = 0 \\ [\,x_l, x_k] = x_\mu x_\nu - x_\nu x_\mu = 0 \end{cases} \qquad (34)$$

with

$$\eta_{\mu\nu} = \begin{cases} 1 & si\; \mu = \nu = 0,1,2,\ldots,D_+ - 1 \\ -1 & si\; \mu = \nu = D_+, D_+ + 2, \ldots, D - 1 \\ 0 & si\; \mu \neq \nu \end{cases} \qquad (35)$$

### 3.4 Multidimensional and relativistic quantum phase space

As in the section 3.2, the definition of the quantum phase space need the identification of the joint momenta-coordinates states which correspond, among other things, to the saturation of the uncertainty relations. Following an approach introduced in [51], these states can also be identified as the simplest states which are naturally associated to the symmetry corresponding to Linear Canonical Transformations (LCTs). The form of the wavefunctions corresponding to these states are "covariants" under the integrals representations of LCTs [33, 51]. These wavefunctions are the multidimensional and relativistic generalization of the wavenfunction in the relation (23) and their generic expression is given in the relation (39) below. For the writing of this generic expression, one should introduce the generalization of the statistical mean values and variance-covariances defined in the relation (20) which are

$$\begin{cases} \langle p_\mu \rangle = \langle \psi | \boldsymbol{p_\mu} | \psi \rangle = \eta_{\mu\nu} \langle \psi | \boldsymbol{p^\nu} | \psi \rangle = \eta_{\mu\nu} \langle p^\nu \rangle \\ \langle x_\mu \rangle = \langle \psi | \boldsymbol{x_\mu} | \psi \rangle = \eta_{\mu\nu} \langle \psi | \boldsymbol{x^\nu} | \psi \rangle = \eta_{\mu\nu} \langle x^\nu \rangle \\ \mathcal{P}_{\mu\nu} = \langle \psi | (\boldsymbol{p_\mu} - \langle p_\mu \rangle)(\boldsymbol{p_\nu} - \langle p_\nu \rangle) | \psi \rangle = \eta_{\mu\lambda} \mathcal{P}_\nu^\lambda \\ \mathcal{X}_{\mu\nu} = \langle \psi | (\boldsymbol{x_\mu} - \langle x_\mu \rangle)(\boldsymbol{x_\nu} - \langle x_\nu \rangle) | \psi \rangle = \eta_{\mu\lambda} \mathcal{X}_\nu^\lambda \\ \varrho_{\mu\nu}^{\ltimes} = \langle \psi | (\boldsymbol{p_\mu} - \langle p_\mu \rangle)(\boldsymbol{x_\nu} - \langle x_\nu \rangle) | \psi \rangle \\ \varrho_{\mu\nu}^{\rtimes} = \langle \psi | (\boldsymbol{x_\nu} - \langle x_\nu \rangle)(\boldsymbol{p_\mu} - \langle p_\mu \rangle) | \psi \rangle \\ \varrho_{\mu\nu} = \frac{1}{2}(\varrho_{\mu\nu}^{\ltimes} + \varrho_{\mu\nu}^{\rtimes}) \end{cases} \quad (36)$$

It is also necessary to introduce the following $\mathcal{B}_{\mu\nu}$ parameters which are generalizations of the parameter $\mathcal{B}_{11}$ defined in the relation (24)

$$\mathcal{B}_{\mu\nu} = \frac{1}{4}[(\hbar)^2 \eta_{\mu\rho} + 2i\hbar \varrho_{\mu\rho}] \widetilde{\mathcal{X}}_\nu^\rho \quad (37)$$

in which $\widetilde{\mathcal{X}}_\nu^\rho$ are the elements of the inverse of the coordinates statistical variance-covariance matrix with elements $\mathcal{X}_\rho^\mu = \eta^{\mu\nu} \mathcal{X}_{\nu\rho}$ i.e. one has the relation (with summation on $\lambda$)

$$\mathcal{X}_\lambda^\mu \widetilde{\mathcal{X}}_\nu^\lambda = \delta_\nu^\mu = \begin{cases} 1 \text{ if } \mu = \nu \\ 0 \text{ if } \mu \neq \nu \end{cases} \Leftrightarrow \mathcal{X}_{\lambda\mu} \widetilde{\mathcal{X}}_\nu^\lambda = \eta_{\mu\nu} \quad (38)$$

One has then for the explicit generic expression of the multidimensional and relativistic generalization of the wavefunction (23)

$$\langle \{x^\mu\} | \{\langle z_\mu \rangle\} \rangle = \langle x | \langle z \rangle \rangle = \frac{e^{-\frac{1}{(\hbar)^2} \mathcal{B}_{\mu\nu}(x^\mu - \langle x^\mu \rangle)(x^\nu - \langle x^\nu \rangle) - \frac{i}{\hbar} \langle p_\mu \rangle x^\mu + iK}}{[(2\pi)^D |det[\mathcal{X}_\nu^\mu]|]^{1/4}} \quad (39)$$

with

- $\mathcal{B}_{\mu\nu}$ the parameters (generally complex numbers) defined in the relation (37)
- $|det[\mathcal{X}_\nu^\mu]|$ the absolute value of the determinant of the matrix with elements $\mathcal{X}_\nu^\mu$

- $K$ a real parameter ($e^{iK}$ is an unitary complex number) which doesn't depend on the coordinates variables $x^\mu$ but may depend on the mean values $\langle p_\mu \rangle$ and $\langle x_\mu \rangle$.

The notations $|\langle x \rangle\rangle$ and $|\langle z \rangle\rangle$ are introduced for the quantum states in the relation (39) to lighten the writing. By using the coordinate representations of the momenta and coordinates, it can be checked that the joint momenta-coordinate states $|\langle z \rangle\rangle$ are eigenstates of the operators $\mathbf{z}_\mu$ defined by the following relation [33, 51]

$$\mathbf{z}_\mu = \mathbf{p}_\mu - \frac{2i}{\hbar} \mathcal{B}_{\mu\nu} \mathbf{x}^\mu \tag{40}$$

The corresponding eigenvalue equation is

$$\mathbf{z}_\mu |\langle z \rangle\rangle = \left( \langle p_\mu \rangle - \frac{2i}{\hbar} \mathcal{B}_{\mu\nu} \langle x^\nu \rangle \right) |\langle z \rangle\rangle = \langle z_\mu \rangle |\langle z \rangle\rangle \tag{41}$$

It can also be verified that the statistical momenta-coordinate variance-covariances corresponding to a state $|\langle z \rangle\rangle$ satisfy the following relation [33]

$$\mathcal{P}_{\mu\nu} = \frac{(\hbar)^2}{4} \widetilde{\mathcal{X}}_{\mu\nu} + \varrho_{\mu\alpha} \widetilde{\mathcal{X}}^{\alpha\beta} \varrho_{\nu\beta} \tag{42}$$

The relations (40) and (41) can be considered as the multidimensional and relativistic generalizations of the relations (25) and (26). The relation (42) corresponds to the expression of the saturation of the uncertainty relation in the multidimensional and relativistic case. The concept of multidimensional and relativistic quantum phase space can be introduced using the joint momenta-coordinates states $|\langle z \rangle\rangle$. This quantum phase space can be defined as the set of all possible values of the pairs of mean values $(\langle p_\mu \rangle, \langle x_\mu \rangle)$ of momenta and coordinates observables $\mathbf{p}_\mu$ and $\mathbf{x}_\mu$ corresponding to the states $|\langle z \rangle\rangle$ for given values of the statistical momenta-coordinates variance-covariances. Equivalently, this multidimensional relativistic phase space can also be defined as the set of possible values of all of the complex variables $\langle z_\mu \rangle = \langle p_\mu \rangle - \frac{2i}{\hbar} \mathcal{B}_{\mu\nu} \langle x^\nu \rangle$ for given values of the all of the $\mathcal{B}_{\mu\nu}$.

From the relations in (36), one can define the momenta-coordinates statistical variance-covariance matrix $\begin{pmatrix} \mathcal{P} & \varrho \\ \varrho^T & \mathcal{X} \end{pmatrix}$ which is a $2D \times 2D$ square matrix. Each of the block elements $\mathcal{P}, \mathcal{X}, \varrho$ and $\varrho^T$ are themselves $D \times D$ square matrix : $\mathcal{P}$ is the momenta statistical variance-covariances matrix with elements $\mathcal{P}_{\mu\nu}$, $\mathcal{X}$ is the momenta statistical variance-covariances matrix with elements $\mathcal{X}_{\mu\nu}$, $\varrho$ is the momenta-coordinates statistical covariance matrix with elements $\varrho_{\mu\nu}$ and $\varrho^T$ is the transpose of $\varrho$. One obtains as multidimensional generalization of the decomposition (27)

$$\begin{pmatrix} \mathcal{P} & \varrho \\ \varrho^T & \mathcal{X} \end{pmatrix} = \begin{pmatrix} \mathscr{b} & 0 \\ 2ac\mathscr{b} & a \end{pmatrix}^T \begin{pmatrix} \eta & 0 \\ 0 & \eta \end{pmatrix} \begin{pmatrix} \mathscr{b} & 0 \\ 2ac\mathscr{b} & a \end{pmatrix} \tag{43}$$

in which $\eta$ is the $D \times D$ square matrix with elements $\eta_{\mu\nu}$ defined in the relation (35) and $a, b$ and $c$ are $D \times D$ square matrices satisfying the following relations

$$\begin{cases} ab = ba = \frac{1}{2}I_D & (I_D \text{ is the } D \times D \text{ identity matrix}) \\ a^T = \eta a \eta & a^\dagger = \eta a^T = a\eta \Leftrightarrow a_\mu^{\nu*} = a^{\mu\nu} \\ b^T = \eta b \eta & b^\dagger = \eta b = b^T \eta \Leftrightarrow b_\mu^{\nu*} = b^{\nu\mu} \\ c^T = 2\eta acb\eta \end{cases} \quad (44)$$

The use of the elements $a_\nu^\mu$ of the matrix $a$ considered in the relations (43) and (44) allows the introduction of the generalizations $\mathbf{z}_\mu$ and $\mathbf{z}_\mu^\dagger$ of the ladder operators defined in the relation (29). One has the following relations for their definitions and commutation relations

$$\begin{cases} \mathbf{z}_\mu = a_\mu^\nu (\mathbf{z}_\nu - \langle z_\nu \rangle) \\ \mathbf{z}_\mu^\dagger = a_\mu^{\nu*}(\mathbf{z}_\nu^\dagger - \langle z_\nu \rangle^*) \end{cases} \quad [\mathbf{z}_\mu, \mathbf{z}_\nu^\dagger] = \delta_{\mu\nu} = \begin{cases} 1 \text{ if } \mu = \nu \\ 0 \text{ if } \mu \neq \nu \end{cases} \quad (45)$$

The use of the ladder operators $\mathbf{z}_\mu$ and $\mathbf{z}_\mu^\dagger$ leads to the introduction of the generalizations of the states $|n_1, \langle z_1 \rangle\rangle$ defined in the relation (30). We may denote these generalizations $|\{n_\mu\}, \langle z_\mu \rangle\rangle = |n, \langle z \rangle\rangle$ (with $n$ the set of the $D$ positive integers $n_0, n_1, \dots, n_{D-1}$). One has for the explicit expression of a state $|n, \langle z \rangle\rangle$

$$|n, \langle z \rangle\rangle = [\prod_{\mu=0}^{D-1} \frac{(\mathbf{z}_\mu^\dagger)^{n_\mu}}{\sqrt{n_\mu!}}]|\langle z \rangle\rangle \quad (46)$$

The states $|n, \langle z \rangle\rangle$ are eigenstates of the quadratic operator $\aleph = \delta^{\mu\nu} \mathbf{z}_\mu^\dagger \mathbf{z}_\nu$. The corresponding eigenvalue equation is

$$\aleph |n, \langle z \rangle\rangle = [\sum_{\mu=0}^{D-1} n_\mu]|n, \langle z \rangle\rangle \quad (47)$$

## 4- Phase space representations of quantum mechanics and microstates

The introduction of the concept of quantum phase space leads explicitly and directly to the definition of the concept of phase space representations of quantum mechanics described in this section. We begin with the simple example of a system with one degree of freedom (one pair of momentum-coordinate observable) and the multidimensional and relativistic generalizations are straightforward.

### 4.1 Phase space representations of states and positive definite distribution functions

One considers a system described with a momentum-coordinate operators pair $(\mathbf{p}_1, \mathbf{x}_1)$. The joint momentum-coordinate states are the states $|\langle z_1 \rangle\rangle$ described in the section 3.1. One can consider, for instance, the states $|\langle z_1 \rangle\rangle$ for which the momentum-coordinate statistical variance-covariance matrix is of the form

$$\begin{pmatrix} \mathcal{P}_{11} & \varrho_{11} \\ \varrho_{11} & \mathcal{X}_{11} \end{pmatrix} = \begin{pmatrix} \mathcal{P}_{11} & 0 \\ 0 & \mathcal{X}_{11} \end{pmatrix} \tag{48}$$

In the general cases, we have $\varrho_{11} \neq 0$ but these general cases can be deduced from the cases corresponding to (48) via Linear Canonical Transformations [33]. One can therefore limit the study to the case of the states $|\langle z_1 \rangle\rangle$ corresponding to the relation (48) without loss of generality. The set of these states is a basis of the states space of the system but this basis is not orthogonal. In fact, one has the following relations [33] (with $h$ the Planck constant and $\boldsymbol{I}$ the identity operator)

$$\begin{cases} \langle\langle z_1\rangle|\langle z'_1\rangle\rangle = e^{-\frac{(\langle p_1\rangle-\langle p'_1\rangle)^2}{8\mathcal{P}_{11}} - \frac{(\langle x_1\rangle-\langle x'_1\rangle)^2}{8\mathcal{X}_{11}} - i\frac{(\langle p_1\rangle-\langle p'_1\rangle)(\langle x_1\rangle+\langle x'_1\rangle)}{2\hbar}} \\ \int |\langle z_1\rangle\rangle\langle\langle z_1\rangle| \frac{d\langle p_1\rangle d\langle x_1\rangle}{h} = \boldsymbol{I} \end{cases} \tag{49}$$

The first equality means that the basis $\{|\langle z_1\rangle\rangle\}$ is not orthogonal. The second equality is a closure-like relation. It can be used to deduce a decomposition of any state $|\psi\rangle$

$$|\psi\rangle = \boldsymbol{I}|\psi\rangle = \int |\langle z_1\rangle\rangle\langle\langle z_1\rangle|\psi\rangle \frac{d\langle p_1\rangle d\langle x_1\rangle}{h} = \int \widetilde{\widetilde{\psi}}(\langle z_1\rangle)|\langle z_1\rangle\rangle \frac{d\langle p_1\rangle d\langle x_1\rangle}{h} \tag{50}$$

The relation (50) defines a phase space representation of the state $|\psi\rangle$. To this representation corresponds the wavefunction $\widetilde{\widetilde{\psi}}(\langle z_1\rangle) = \langle\langle z_1\rangle|\psi\rangle$ that one may call wavefunction in a phase space representation or a phase space wavefunction. As the basic states corresponding to this representation are the eigenstates of the operator $\boldsymbol{z}_1$ defined in the relation (25), one may also call this representation the $\boldsymbol{z}$ representation.

Another representation that is linked to the concept of quantum phase spaces is the one which correspond to the basic states $|n_1, \langle z_1\rangle\rangle$ for a fixed value of $\langle z_1\rangle$. In fact, one can show that, for each fixed value of $\langle z_1\rangle$, the ket family $\{|n_1, \langle z_1\rangle\rangle\}$ is an orthonormal basis of the states space of the system [33]. More explicitly, one has the following relations

$$\begin{cases} \langle n_1, \langle z_1\rangle|n'_1, \langle z_1\rangle\rangle = \delta_{n_1 n'_1} = \begin{cases} 1 \ si \ n_1 = n'_1 \\ 0 \ si \ n_1 \neq n'_1 \end{cases} \\ \sum_{n_1} |n_1, \langle z_1\rangle\rangle\langle n_1, \langle z_1\rangle| = \boldsymbol{I} \end{cases} \tag{51}$$

The decomposition in the basis $\{|n_1, \langle z_1\rangle\rangle\}$ of any state $|\psi\rangle$ is as follows

$$|\psi\rangle = \boldsymbol{I}|\psi\rangle = \sum_{n_1} |n_1, \langle z_1\rangle\rangle\langle n_1, \langle z_1\rangle|\psi\rangle = \sum_{n_1} \widetilde{\widetilde{\psi}}(n_1, \langle z_1\rangle)|n_1, \langle z_1\rangle\rangle \tag{52}$$

As the states $|n_1, \langle z_1\rangle\rangle$ corresponding to the representation (52) are, according to the relation (31), eigenstates of the operator $\aleph$, we may call this representation the $\aleph$ representation. Following the relation (52), the wavefunction corresponding to the state $|\psi\rangle$ in this representation is the function $\widetilde{\widetilde{\psi}}(n_1, \langle z_1\rangle) = \langle n_1, \langle z_1\rangle|\psi\rangle$. It can be remarked that the value on $\langle z_1 \rangle$ of the wavefunction $\widetilde{\psi}(\langle z_1 \rangle) = \langle\langle z_1|\psi\rangle$ corresponding to the **z** representation is equal to the value of $\widetilde{\psi}(n_1, \langle z_1 \rangle)$ for $n_1 = 0$ i.e. $\widetilde{\psi}(0, \langle z_1 \rangle) = \widetilde{\psi}(\langle z_1 \rangle) = \langle\langle z_1|\psi\rangle$.

As the functions $\widetilde{\psi}(\langle z_1 \rangle)$ and $\widetilde{\psi}(n_1, \langle z_1 \rangle)$ are wavefunctions, the squares $\left|\widetilde{\psi}(\langle z_1 \rangle)\right|^2 = |\langle\langle z_1|\psi\rangle|^2$ and $\left|\widetilde{\psi}(n_1, \langle z_1 \rangle)\right|^2 = |\langle n_1, \langle z_1 \rangle|\psi\rangle|^2$ of their modulus have probabilistic interpretations. More precisely:

- $\left|\widetilde{\psi}(\langle z_1 \rangle)\right|^2 = |\langle\langle z_1|\psi\rangle|^2$ is the density of probability to find the system in a state $|\langle z_1 \rangle\rangle$ knowing that its state is exactly $|\psi\rangle$. In other words, $\left|\widetilde{\psi}(\langle z_1 \rangle)\right|^2$ is the density of probability of localization in the quantum phase space $\{\langle z_1 \rangle\}$ if the quantum state of the system is $|\psi\rangle$. The corresponding normalization relation is [33]

$$\int \left|\widetilde{\psi}(\langle z_1 \rangle)\right|^2 \frac{d\langle p_1\rangle d\langle x_1\rangle}{h} = \int |\langle\langle z_1|\psi\rangle|^2 \frac{d\langle p_1\rangle d\langle x_1\rangle}{h} = 1 \qquad (53)$$

- $\left|\widetilde{\psi}(n_1, \langle z_1 \rangle)\right|^2 = |\langle n_1, \langle z_1 \rangle|\psi\rangle|^2$ is, for a fixed value of $\langle z_1 \rangle$, the probability to find the system in a state $|n_1, \langle z_1 \rangle\rangle$ knowing that its state is exactly $|\psi\rangle$. The normalization relation is

$$\sum_{n_1} \left|\widetilde{\psi}(n_1, \langle z_1 \rangle)\right|^2 = \sum_{n_1} |\langle n_1, \langle z_1 \rangle|\psi\rangle|^2 = 1 \qquad (54)$$

The density of probability $\left|\widetilde{\psi}(\langle z_1 \rangle)\right|^2 = |\langle\langle z_1|\psi\rangle|^2 = \langle\langle z_1|\psi\rangle\langle\psi|\langle z_1\rangle\rangle$ corresponds to a system in a pure state $|\psi\rangle$ which corresponds to the density operator $\boldsymbol{\rho} = |\psi\rangle\langle\psi|$. But one may consider a more general expression of the density of probability of localization in the quantum phase space for a system which is in a mixed state described by a density operator having the form given in the relation (15). The expression of this general density of probability, which is a positive definite quantum phase space distribution function, is

$$\rho(\langle z_1 \rangle) = \langle\langle z_1|\boldsymbol{\rho}|\langle z_1\rangle\rangle = \sum_I p_I \langle\langle z_1|\boldsymbol{\rho_I}|\langle z_1\rangle\rangle = \sum_I p_I |\langle\langle z_1|\psi_I\rangle|^2 \qquad (55)$$

The function $\rho(\langle z_1 \rangle)$ defined in the relation (55) is exactly the analogous, in the case of a quantum phase space, of the Wigner distribution which is defined with a classical phase space. But unlike this Wigner distribution, $\rho(\langle z_1 \rangle)$ is a positive definite phase space distribution function. The corresponding normalization relation is

$$\int \rho(\langle z_1 \rangle) \frac{d\langle p_1\rangle d\langle x_1\rangle}{h} = \sum_I p_I \int |\langle\langle z_1|\psi_I\rangle|^2 \frac{d\langle p_1\rangle d\langle x_1\rangle}{h} = \sum_I p_I = 1 \qquad (56)$$

It is straightforward to have the multidimensional and relativistic generalization of the relations defining the phase space representations discussed previously. One has, for instance, for the generalizations of the closure relations and the decompositions of any state $|\psi\rangle$:

$$\int |\langle z\rangle\rangle\langle\langle z\rangle| \prod_{\mu=0}^{D-1} \frac{d\langle p_\mu\rangle d\langle x_\mu\rangle}{(h)^D} = I \tag{57}$$

$$|\psi\rangle = I|\psi\rangle = \int |\langle z\rangle\rangle\langle\langle z\rangle|\psi\rangle \prod_{\mu=0}^{D-1} \frac{d\langle p_\mu\rangle d\langle x_\mu\rangle}{(h)^D} = \int \widetilde{\psi}(\langle z\rangle)|\langle z\rangle\rangle \prod_{\mu=0}^{D-1} \frac{d\langle p_\mu\rangle d\langle x_\mu\rangle}{(h)^D} \tag{58}$$

$$\sum_{n_1,n_2,\ldots,n_{D-1}} |n,\langle z\rangle\rangle\langle n,\langle z\rangle| = I \tag{59}$$

$$|\psi\rangle = I|\psi\rangle = \sum_{n_1,n_2,\ldots,n_{D-1}} |n,\langle z\rangle\rangle\langle n,\langle z\rangle|\psi\rangle = \sum_{n_1,n_2,\ldots,n_{D-1}} \widetilde{\psi}(n,\langle z\rangle)|n,\langle z\rangle\rangle \tag{60}$$

The multidimensional and relativistic generalizations of the relations (55) and (56) corresponding to the positive definite phase space distribution can also be established. These relations are

$$\rho(\langle z\rangle) = \rho(\langle z_0\rangle,\langle z_1\rangle,\ldots,\langle z_{D-1}\rangle) = \langle\langle z\rangle|\rho|\langle z\rangle\rangle \tag{61}$$

$$\int \rho(\langle z\rangle) \prod_{\mu=0}^{D-1} \frac{d\langle p_\mu\rangle d\langle x_\mu\rangle}{(h)^D} = 1 \tag{62}$$

### 4.2 Phase space representations of operators

Let us firstly consider the case of a system with one degree of freedom i.e. described by one pair $(\boldsymbol{p}_1, \boldsymbol{x}_1)$ of momentum-coordinate observables. There are two ways to represent a quantum operator $\boldsymbol{A}$ in the framework of the $\boldsymbol{z}$ representation :

- by using a decomposition in the basis $\{|\langle z_1\rangle\rangle\}$ which is based on the relation (49)

$$\boldsymbol{A} = I\boldsymbol{A}I = \iint\iint |\langle z_1\rangle\rangle\langle\langle z_1\rangle|\boldsymbol{A}|\langle z_1\rangle'\rangle\langle\langle z_1\rangle'| \frac{d\langle p_1\rangle d\langle x_1\rangle}{h} \frac{d\langle p_1\rangle' d\langle x_1\rangle'}{h} \tag{63}$$

The function $\langle\langle z_1\rangle|\boldsymbol{A}|\langle z_1\rangle'\rangle = A(\langle z_1\rangle,\langle z_1\rangle')$ which represents the operator $\boldsymbol{A}$ according to the relation (63) is a function of the two variables $\langle z_1\rangle$ and $\langle z_1\rangle'$ and is a kind of "continuous matrix". It can be considered as a quantum generalization of the functions which represent observables in classical phase space in the framework of classical Hamiltonian mechanics.

- with a representative operator $\widetilde{\boldsymbol{A}}$ which acts on the set $\{\widetilde{\psi}\}$ of phase space wavefunctions $\widetilde{\psi}(\langle z_1\rangle)$ and which is defined by the relation

$$\langle\langle z_1\rangle|\boldsymbol{A}|\psi\rangle = \widetilde{\boldsymbol{A}}\,[\langle\langle z_1\rangle|\psi\rangle] = \widetilde{\boldsymbol{A}}\,[\widetilde{\psi}(\langle z_1\rangle)] \tag{64}$$

The relation (64) can be for instance applied to the case of the momentum and coordinate observables $\boldsymbol{p_1}$ and $\boldsymbol{x_1}$. One should find the operators $\widetilde{\boldsymbol{p}}_1$ and $\widetilde{\boldsymbol{x}}_1$ which satisfy the relations

$$\begin{cases} \langle\langle z_1\rangle|\boldsymbol{p_1}|\psi\rangle = \widetilde{\boldsymbol{p}}_1\left[\langle\langle z_1\rangle|\psi\rangle\right] = \widetilde{\boldsymbol{p}}_1\left[\widetilde{\psi}\left(\langle z_1\rangle\right)\right] \\ \langle\langle z_1\rangle|\boldsymbol{x_1}|\psi\rangle = \widetilde{\boldsymbol{x}}_1\left[\langle\langle z_1\rangle|\psi\rangle\right] = \widetilde{\boldsymbol{x}}_1\left[\widetilde{\psi}\left(\langle z_1\rangle\right)\right] \end{cases} \quad (65)$$

Given the relations (2) and (3), one can deduce the following relations

$$\begin{cases} \langle\langle z_1\rangle|\boldsymbol{p_1}|\psi\rangle = \int \langle\langle z_1\rangle|p_1\rangle \langle p_1|\boldsymbol{p_1}|\psi\rangle dp_1 = \int p_1 \langle\langle z_1\rangle|\langle p_1\rangle\rangle \widetilde{\psi}(\langle p_1\rangle) dp_1 \\ \langle\langle z_1\rangle|\boldsymbol{x_1}|\psi\rangle = \int \langle\langle z_1\rangle|x_1\rangle \langle x_1|\boldsymbol{x_1}|\psi\rangle dx_1 = \int x_1 \langle\langle z_1\rangle|x_1\rangle \psi(\langle x_1\rangle) dx_1 \end{cases} \quad (66)$$

The expressions of the functions $\langle\langle z_1\rangle|\langle x_1\rangle\rangle$ and $\langle\langle z_1\rangle|\langle p_1\rangle\rangle$ can be deduced from the relations (4) and (23). One finds

$$\begin{cases} \langle x_1|\langle z_1\rangle\rangle = \left(\dfrac{1}{2\pi\mathcal{X}_{11}}\right)^{1/4} e^{-\frac{B_{11}}{\hbar^2}(x^1-\langle x^1\rangle)^2 - \frac{i}{\hbar}\langle p_1\rangle x^1 + iK} \\ \langle\langle z_1\rangle|x_1\rangle = (\langle x_1|\langle z_1\rangle\rangle)^* = \left(\dfrac{1}{2\pi\mathcal{X}_{11}}\right)^{1/4} e^{-\frac{B_{11}}{\hbar^2}(x^1-\langle x^1\rangle)^2 + \frac{i}{\hbar}\langle p_1\rangle x^1 - iK} \\ \langle\langle z_1\rangle|p_1\rangle = \int \langle\langle z_1\rangle|x_1\rangle\langle x_1|\langle p_1\rangle\rangle\, dx_1 = \left(\dfrac{1}{2\pi\mathcal{P}_{11}}\right)^{1/4} e^{-\frac{\mathcal{A}_{11}}{\hbar^2}(p_1-\langle p_1\rangle)^2 + \frac{i}{\hbar}(p_1-\langle p_1\rangle)x^1 - iK} \end{cases} \quad (67)$$

One deduces from the relations (66) and (67) the following relations

$$\begin{cases} \langle\langle z_1\rangle|\boldsymbol{p_1}|\psi\rangle = \left(\dfrac{1}{2\pi\mathcal{P}_{11}}\right)^{1/4} \int [p_1\,\widetilde{\psi}(\langle p_1\rangle)] e^{-\frac{\mathcal{A}_{11}}{\hbar^2}(p_1-\langle p_1\rangle)^2 + \frac{i}{\hbar}(p_1-\langle p_1\rangle)x^1 - iK} \\ \langle\langle z_1\rangle|\boldsymbol{x_1}|\psi\rangle = \left(\dfrac{1}{2\pi\mathcal{X}_{11}}\right)^{1/4} \int [x_1\,\psi(\langle x_1\rangle)] e^{-\frac{B_{11}}{\hbar^2}(x^1-\langle x^1\rangle)^2 + \frac{i}{\hbar}\langle p_1\rangle x^1 - iK} \end{cases} \quad (68)$$

Then, using integrations by parts and the relations in (65), (66), (67) and (68), one can finally deduce the expressions of the operators $\widetilde{\boldsymbol{p}}_1$ and $\widetilde{\boldsymbol{x}}_1$ which are the phase space representations of the momentum and coordinates operators

$$\begin{cases} \widetilde{\boldsymbol{p}}_1 = -\widetilde{p^1} = i\hbar\dfrac{\partial}{\partial\langle x^1\rangle} + \langle p_1\rangle + \hbar\dfrac{\partial K}{\partial\langle x^1\rangle} \\ \widetilde{\boldsymbol{x}}_1 = -\widetilde{x^1} = -i\hbar\dfrac{\partial}{\partial\langle p^1\rangle} - \hbar\dfrac{\partial K}{\partial\langle p^1\rangle} \end{cases} \quad [\widetilde{\boldsymbol{p}}_1, \widetilde{\boldsymbol{x}}_1] = -i\hbar \quad (69)$$

The parameters $K$ in the relations (69) is the same as in the expression of the wavefunction $\widetilde{\psi}(\langle z_1\rangle)$ in the relation (23). According to (69), the explicit expressions of the operators $\widetilde{\boldsymbol{p}}_1$ and $\widetilde{\boldsymbol{x}}_1$ depend on the expression that is chosen for $K$ but the CCR does not depend on $K$.

For the case of the ℵ representation associated to the basic states $|n_1, \langle z_1 \rangle\rangle$ for a fixed value of $\langle z_1 \rangle$, one has matrices representations of the operators. More explicitly, an operator $A$ is represented in this framework by a matrix $[A]$ with elements $A_{n_1'}^{n_1}$ so that one has the relations

$$\begin{cases} A = \sum_{n_1} \sum_{n_1'} A_{n_1'}^{n_1} |n_1, \langle z_1 \rangle\rangle \langle n_1', \langle z_1 \rangle| \\ A_{n_1'}^{n_1} = \langle n_1, \langle z_1 \rangle | A | n_1', \langle z_1 \rangle \rangle \end{cases} \quad (70)$$

It is straightforward to deduce the multidimensional and relativistic generalizations of the relations (63), (64), (65), (69) and (70). On has, for instance, for the generalization of the relation (69) which correspond to a $z$ representation of the momenta and coordinates operators

$$\begin{cases} \widetilde{\boldsymbol{p}}_\mu = i\hbar \dfrac{\partial}{\partial \langle x^\mu \rangle} + \langle p_\mu \rangle + \hbar \dfrac{\partial K}{\partial \langle x^\mu \rangle} \\ \widetilde{\boldsymbol{x}}_\mu = -i\hbar \dfrac{\partial}{\partial \langle p^\mu \rangle} - \hbar \dfrac{\partial K}{\partial \langle p^\mu \rangle} \end{cases} \quad [\widetilde{\boldsymbol{p}}_\mu, \widetilde{\boldsymbol{x}}_\nu] = i\hbar \eta_{\mu\nu} \quad (71)$$

In which $\eta_{\mu\nu}$ and $K$ are the same as in the relations (35) and (39). At this level, no particular constraints can be identified in the choice of the expressions of $K$. Various choice can be considered, one has for instance:

- For $K = 0$

$$\begin{cases} \widetilde{\boldsymbol{p}}_\mu = i\hbar \dfrac{\partial}{\partial \langle x^\mu \rangle} + \langle p_\mu \rangle \\ \widetilde{\boldsymbol{x}}_\mu = -i\hbar \dfrac{\partial}{\partial \langle p^\mu \rangle} \end{cases} \quad [\widetilde{\boldsymbol{p}}_\mu, \widetilde{\boldsymbol{x}}_\nu] = i\hbar \eta_{\mu\nu} \quad (72)$$

- For $K = -\dfrac{1}{\hbar} \langle p_\mu \rangle \langle x^\mu \rangle$ (with summation on $\mu$)

$$\begin{cases} \widetilde{\boldsymbol{p}}_\mu = i\hbar \dfrac{\partial}{\partial \langle x^\mu \rangle} \\ \widetilde{\boldsymbol{x}}_\mu = -i\hbar \dfrac{\partial}{\partial \langle p^\mu \rangle} + \langle x_\mu \rangle \end{cases} \quad [\widetilde{\boldsymbol{p}}_\mu, \widetilde{\boldsymbol{x}}_\nu] = i\hbar \eta_{\mu\nu} \quad (73)$$

- For $K = -\dfrac{1}{2\hbar} \langle p_\mu \rangle \langle x^\mu \rangle$ (with summation on $\mu$)

$$\begin{cases} \widetilde{\boldsymbol{p}}_\mu = i\hbar \dfrac{\partial}{\partial \langle x^\mu \rangle} + \dfrac{1}{2} \langle p_\mu \rangle \\ \widetilde{\boldsymbol{x}}_\mu = -i\hbar \dfrac{\partial}{\partial \langle p^\mu \rangle} + \dfrac{1}{2} \langle x_\mu \rangle \end{cases} \quad [\widetilde{\boldsymbol{p}}_\mu, \widetilde{\boldsymbol{x}}_\nu] = i\hbar \eta_{\mu\nu} \quad (74)$$

It can be checked, in particular, that the CCRs don't depend on the expression of $K$.

## 4.3 Rigorous descriptions of microstates

As already said in the section 2.3, it is common to identify a microstate with a quantum pure state in the framework of quantum statistical mechanics. However, as it was also highlighted, even the microstates considered in "classical statistical mechanics" should have a finite phase space hypervolume and our goal in this section is to highlight the fact that "any microstate" should be explicitly and always identified as a quantum pure state. The contents of this section also highlights the existence of fundamental links between quantum states, quantum phase space and statistical mechanics.

For the case of a particle with one degree of freedom, the "elementary surface" in phase space i.e. the surface corresponding to a statistical mechanic microstate is equal to the Planck constant $h$. This fact can be directly related to the normalization relation (53) by identifying a microstate with a quantum state $|\psi\rangle$ and the surface occupied by this microstate with the integral on the whole phase space of $\left|\widetilde{\psi}(\langle z_1\rangle)\right|^2 = |\langle\langle z_1\rangle|\psi\rangle|^2$ which is also exactly equal to $h$. In fact, the following relation can be deduced easily from the relation (53)

$$\int \left|\widetilde{\psi}(\langle z_1\rangle)\right|^2 d\langle p_1\rangle d\langle x_1\rangle = \int |\langle\langle z_1\rangle|\psi\rangle|^2 d\langle p_1\rangle d\langle x_1\rangle = h \quad (75)$$

In the framework of this point of view, a microstate is not a point in phase space and it doesn't even correspond to a naïve vision of a finite region of phase space with surface equal to $h$ but it "spreads" on the whole quantum phase space and to this "spreadness" corresponds at each point of the phase space a weight equal to $\left|\widetilde{\psi}(\langle z_1\rangle)\right|^2 = |\langle\langle z_1\rangle|\psi\rangle|^2$. It is the integral of $|\langle\langle z_1\rangle|\psi\rangle|^2$ on the "whole phase space" that is equal to $h$ as shown in the relation (75). The multidimensional and relativistic generalization of these results is straightforward. One has for instance for the generalization of the relation (75)

$$\int \left|\widetilde{\psi}(\langle z_0\rangle,\langle z_1\rangle,\ldots\langle z_{D-1}\rangle)\right|^2 \prod_{\mu=0}^{D-1} d\langle p_\mu\rangle d\langle x_\mu\rangle = \int |\langle\langle z\rangle|\psi\rangle|^2 \prod_{\mu=0}^{D-1} d\langle p_\mu\rangle d\langle x_\mu\rangle = (h)^D \quad (76)$$

The relation (76) means that a microstate of the system can be identified with a quantum state $|\psi\rangle$ described by the phase space wavefunction $\widetilde{\psi}(\langle z_0\rangle,\langle z_1\rangle,\ldots,\langle z_{D-1}\rangle) = \langle\langle z\rangle|\psi\rangle$ and the hypervolume in the quantum phase space corresponding to this microstate is equal to $(h)^D$. While having a finite hypervolume, the microstate doesn't correspond to a finite region of the relativistic quantum phase space but is spread out on the whole phase space and to this "spreadness" correspond at each point $\langle z\rangle$ of this phase space a weight equal to $\left|\widetilde{\psi}(\langle z_0\rangle,\langle z_1\rangle,\ldots\langle z_{D-1}\rangle)\right|^2 = |\langle\langle z\rangle|\psi\rangle|^2$.

## 5- Discussions and conclusions.

The wave-particle duality which was considered as the answer to the old question related to the nature of light led to the development of quantum physics. It was shown through this work that it permits also to bring more clarity in the understanding of some important concepts introduced previously in classical physics, more precisely the concepts of phase space and microstates. In fact, these concepts were firstly introduced in classical Hamiltonian mechanics and classical statistical mechanics before the discovery of the wave-particle duality and the uncertainty principle. However, it is highlighted by this works that a more correct understanding of their meanings cannot be done without quantum mechanics. According to the relations (75) and (76), the existence of the nonzero value for the phase space hypervolume of a microstate, for instance, cannot be explained explicitly without the concepts of quantum phase space and phase space representations of quantum mechanics. Following the section 4.1, these concepts permit also, among other things, the obtaining of positive definite quantum phase space distribution functions. The structure of a quantum phase space is shaped by the statistical momenta-coordinates variance-covariance which can be related directly with thermodynamic variables as shown in the references [33-34]. So the concept of quantum phase space provides a way for describing the natural relations between quantum mechanics, phase space, statistical mechanics and thermodynamics.

According to the description given in the section 3 and 4, the concepts of quantum phase space and phase space representation can be considered as a kind of "joint energy-momenta-coordinates representation". But on one hand according to the Planck-Einstein-De Broglie relations in (1), energy and momenta are directly proportional to frequency and wave vector so they can be considered as highlighting the wave nature of a particle. And on the other hand, the concepts of coordinates can be considered as highlighting the corpuscle nature of a particle. It follows that the phase space representation, as a "joint energy-momenta-coordinates representation", can be considered as providing, from a physico-mathematical point of view, a deeper description of the wave-particle duality.

It can also be expected that the results described in the present work can help in the study and resolutions of some current issues related to foundational problems of quantum mechanics like quantum decoherence and the measurement problem. As briefly discussed in the section 3.2, it is for instance considered in the study of quantum decoherence that the environment-induced superselection of pointer states, can be considered as a results of a natural "continuous measure" through collisions and scattering processes of a given particle with the surrounding particles belonging to its environment [12]. But the physical observables that are continuously, mutually and naturally "measured" through these processes are the particle momenta and coordinates, so the real pointer states in this case should correspond to some "joint momenta-coordinates" states i.e. the states that are associated to the concept of quantum phase space.

The fact that the results described in the present work include also multidimensional and relativistic cases may lead to more interesting applications of them. In fact, these results can be for instance used directly in the study of decoherence and the measurement problem in the framework of a relativistic approach. This approach will permit, among other things, to establish explicit relations between these problems and results from relativistic particle physics in which particles nature and properties are directly related to the symmetry group of the physical theory which describes them [33, 51,66-67]. Establishing and studying direct relations between foundational problems of quantum mechanics and particle physics using a framework based on the concept of relativistic quantum phase space could lead to interesting breakthroughs and results.